\documentclass[sigconf]{acmart}

\AtBeginDocument{%
  }

\setcopyright{acmlicensed}
\copyrightyear{2018}
\acmYear{2018}
\acmDOI{XXXXXXX.XXXXXXX}

\acmConference[Conference acronym 'XX]{Make sure to enter the correct
  conference title from your rights confirmation emai}{June 03--05,
  2018}{Woodstock, NY}
\acmISBN{978-1-4503-XXXX-X/18/06}

\usepackage{subfig}
\begin{document}

\title{A Comparative Study on Recommendation
Algorithms: Online and Offline Evaluations on a Large-scale Recommender System}
\title{Online and Offline Evaluations of Collaborative Filtering and Content Based Recommender Systems}
\author{Ali Elahi}
\email{aelahi6@uic.edu}
\orcid{}
\affiliation{%
  \institution{University of Illinois at Chicago}
  \city{Chicago}
  \state{Illinois}
  \country{USA}
}

\author{Armin Zirak}
\email{aarmin.zirak@ucalgary.ca}
\orcid{}
\affiliation{%
  \institution{University of Calgary}
  \state{Calgary}
  \country{Canada}
}

\settopmatter{printacmref=false} 
\renewcommand\footnotetextcopyrightpermission[1]{} 


\begin{abstract}
    Recommender systems are widely used AI applications designed to help users efficiently discover relevant items. The effectiveness of such systems is tied to the satisfaction of both users and providers. However, user satisfaction is complex and cannot be easily framed mathematically using information retrieval and accuracy metrics. While many  studies evaluate accuracy through offline tests, a growing number of researchers argue that online evaluation methods such as A/B testing are better suited for this purpose.
    
    We have employed a variety of algorithms on different types of datasets divergent in size and subject, producing recommendations in various platforms, including media streaming services, digital publishing websites, e-commerce systems, and news broadcasting networks. Notably, our target websites and datasets are in Persian (Farsi) language.
    
    This study provides a comparative analysis of a large-scale recommender system that has been operating for the past year across about 70 websites in Iran, processing roughly 300 requests per second collectively. The system employs user-based and item-based recommendations using content-based, collaborative filtering, trend-based methods, and hybrid approaches. Through both offline and online evaluations, we aim to identify where these algorithms perform most efficiently and determine the best method for our specific needs, considering the dataset and system scale.
    
    Our methods of evaluation include manual evaluation, offline tests including accuracy and ranking metrics like hit-rate@k and nDCG, and online tests consisting of click-through rate (CTR). Additionally we analyzed and proposed methods to address cold-start and popularity bias.
\end{abstract}



\maketitle

\section{Introduction}
Recommender Systems (RS) are one of the most common and significant services provided by information systems. Music and media streaming platforms, e-commerce, employment websites, and online news publishers all employ RS to display content more efficiently to their users \cite{14, 23, 15}. RS filter streams of information to present the user with related or personalized content. This filtering process is either focused on item similarity or users' previous views, the first being item-based and the latter being user-based.

More traditional methods, use CB and CF algorithms analyzing the contextual similarity and feedback datasets. Recently, reinforcement and deep learning-based methods are being utilized in the state-of-the-studies. A gap, however, is apparent between the practical and the academic environments, which this paper aims to cover via the evaluation of the algorithms in the practical environment.

Parallel to ``Recommendation with a Purpose'' \cite{16}, our ultimate objective is to enhance the efficiency of methods in real-world environments. A key distinction between academic and practical settings lies in the evaluation approach. While offline tests, often used in academic research, focus on metrics like accuracy and diversity, online evaluations measure performance through metrics like click-through rate (CTR) and time spent on the platform (PPS). Rossetti et al. \cite{24} have shown that offline metrics often fail to accurately predict algorithm performance in real-world scenarios. Another factor contributing to the gap between academic research and practical applications is the limited variety and scale of datasets used in academic studies. In practical environments, the attributes of recommended items and user behavior are dynamic across different contexts, leading to unpredictable feedback.

A/B testing is a common online testing method. In A/B testing, users are presented with different sets of recommendations, which can result in some groups receiving superior suggestions while others receive inferior ones. The two groups are either exposed to the same algorithm with varying hyper-parameters or two distinct algorithms, typically with one key variant distinguishing the two groups. Online evaluations must be conducted continuously, with performance comparisons assessed over time, as the effectiveness of the recommendation system (RS) is dynamic and can fluctuate due to factors like users' growing trust in the system or cold-start challenges. \cite{6}.

This study was conducted in a company that provides RS as a service. It has been providing over 30 million users within the past year; processing with a load of 300 requests per second and maintains a clientele of approximately 70 websites. The service delivers user-based and item-based recommendation systems engaging content-based (CB), collaborative filtering (CF), trend-based, and some hybrids.

We tested our RS on a range of platforms, including e-commerce sites, media streaming services, news broadcasting websites, and digital publishing networks. In our analysis, we accounted for the differences in type and scale of data across these websites, recognizing that various data sets can yield differing performance levels with different algorithms. We thoroughly explored multiple algorithms, made comparisons, and considered the contextual factors that influence their effectiveness.

Our offline analysis includes ranking and accuracy metrics such as hit-rate@k and, nDCG, mainly used for hyper-parameter tuning. CTR is used as our online metric.

\begin{figure*}
\centering
  
  \includegraphics[width=450pt]{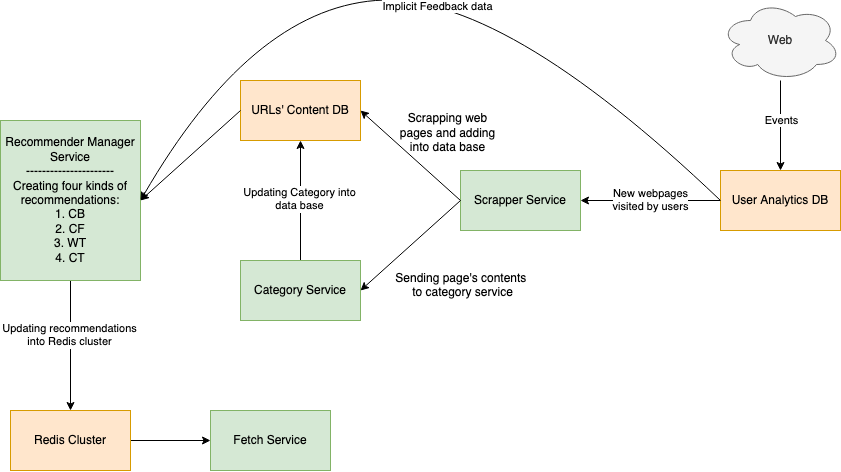}

\caption{The RS architecture is shown above; The green boxes are the services and the orange boxes are data-bases.}
\label{flow}
\end{figure*}

\vspace{-3.5mm}

\section{Literature Review}
\subsection{Evaluation of RS}

Traditionally, the primary metrics for evaluating recommendation systems focus on accuracy, such as Mean Average Precision (MAP), Normalized Discounted Cumulative Gain (NDCG), and Area Under the Curve (AUC). However, relying solely on accuracy can be insufficient and may even hinder the performance of the recommendation system \cite{21}. Users tend to prefer recommendations that are exciting, surprising, novel, and diverse. Novelty refers to how unique an item appears to the user, while diversity is achieved by ensuring that the recommended items differ from one another \cite{8}.

Numerous studies on RS evaluation have established that offline tests, when used in isolation, fail to accurately reflect a system's efficiency \cite{13, 7}. In response to this limitation, several researchers have explored ways to enhance offline testing methodologies. Some studies \cite{26, 28} have integrated diversity and novelty metrics alongside traditional accuracy measurements to provide a more comprehensive assessment.

Studies propose that RS online evaluations, should be done frequently \cite{23, 6}. The performances of a RS may change over time in an online environment. Users often need time to develop trust in the system through their interactions.

Jeunen et al. \cite{17} proposed evaluating CF methods by using the most recent segment of the dataset rather than a random subset. This led to the introduction of SW-eval (sliding window evaluation), which is presented as a superior alternative to Leave-One-Out Cross-Validation (LOOCV), offering improved insights into the performance of CF techniques.

Another evaluation method is using questionnaires, offering insights into user satisfaction and preferences \cite{24}. Other research has focused on correlating offline and online tests to predict CTR using offline metrics, demonstrating the relationship between these evaluation methods \cite{22, 25, 19, 20}. Furthermore, the concept of viewing the RS as a black box has been proposed, facilitating a better understanding of model behavior and improving evaluation methods that aid in model selection \cite{11}. These approaches collectively enhance the evaluation and implementation of RS.

Jannach et al. \cite{16} has evaluated RS both users and providers perspective. Users seek to discover new items and alternatives within their areas of interest, while also wanting their needs to be met. On the other hand, providers aim to generate new demands among users, introduce new services, enhance user retention, and ultimately increase revenue. It is essential to have metrics in place to assess whether these needs are being satisfied. Jannach et al. \cite{16} introduces various scenarios that users may encounter, which can be leveraged to improve the accuracy of our recommendations. A user might be casually browsing, looking for similar items, or specifically searching for popular products.

\subsection{Challenges in Recommender Systems}

The effectiveness of recommendation algorithms are context dependent, as they each come with their own set of benefits and drawbacks. CF algorithms often exhibit popularity biases, frequently recommending popular items over others. This can be influenced by the website's landing page, which may highlight certain items more prominently, leading to an unfairly higher number of views for those items. Implicit feedback datasets \cite{9} are particularly affected by this popularity bias, as they are based solely on clicks and views, whereas explicit feedback datasets rely on user rankings and like/dislike buttons. Items categorized as ``long tail,'' which can be more effective drivers of growth for providers, are recommended significantly less than a select few popular items \cite{1, 18}. Metrics such as the Gini index and group average popularity are used to illustrate these complexities. Studies have proposed filtering methods or re-ranking methods \cite{2, 3} to mitigate the effect of item exposure.

CF algorithms also tend to have lower fill rates and coverage. The fill rate refers to the proportion of users or items for which the RS can generate recommendations, while coverage indicates the percentage of items that the system is able to recommend. One factor contributing to the lower fill rate in CF is the cold start problem, which occurs when there is insufficient data about certain users or items. This issue is particularly prevalent on websites with high item churn, such as news sites that publish hundreds of articles daily. The paper titled “When Collaborative Filtering’s Data Is Not Enough” \cite{12} discusses this challenge in detail. To address the cold start problem, state-of-the-art studies like \cite{5} often utilize hybrid methods.

\section{Methodology}
Our RS provides service to 70 websites. We track the pages each user views and use this data to build an implicit feedback matrix. Contents of the pages are also crawled for the CB algorithm. A variety of algorithms have been implemented, which will be introduced at length.

\subsection{Algorithms Overview}

\paragraph{Website-Trend} Trend-based recommendation is an item-to-item algorithm that suggests pages currently trending on a website. 

\paragraph{Content-based} For content-based (CB) algorithms, a word2vec embedding is generated for each item using its title and the first L words of its text. The word2vec model, specifically trained in Farsi on a 50-gigabyte corpus with diverse contexts using the Gensim library, provides these embeddings. This mapping allows items to be represented as vectors, and a nearest neighbor algorithm is then applied to recommend similar pages based on their proximity in the embedding space.

\paragraph{Category-Trend} Using Latent Semantic Analysis (LSA), we created mappings for the pages. We then applied the DBSCAN clustering method to group pages, and used the Bayesian Information Criterion (BIC) to determine the optimal number of categories. For recommending items related to a specific item I in category C, popular items from the same category are randomly selected.

\paragraph{Collaborative-Filtering} CF algorithm utilize matrix factorization to generate item-to-item and user-to-item recommendations. We used the Alternating Least Squares (ALS) algorithm \cite{5197422} implemented by the Implicit library. Section \ref{als} will analyses the performance of this algorithm. In our CF method, we used implicit feedback matrix \cite{9}, populated by page view counts. This approach generates embeddings for both items and users, with item-based recommendations determined by identifying each item’s nearest neighbors in the embedding space. To approximate the implicit feedback matrix, the user and item matrices are multiplied, enabling user-based recommendations by suggesting unseen items with the highest predicted values.

\subsection{System Architecture}
Figure \ref{flow} illustrates the architecture of our RS. The User Analytics database stores user browsing data and creates the implicit feedback dataset, while the Scraper service collects item URLs viewed by users that haven't yet been saved, populating the URL-content database. The Recommender Manager service applies various algorithms using this data to generate item- and user-based recommendations, loading the results into the REDIS dataset. As a page loads, the Fetch service retrieves appropriate recommendations from REDIS. For new pages in a domain, the Recommender Manager periodically updates REDIS, with the frequency depending on the site’s item turnover. High-churn websites, like news sites, receive updates more frequently. Additionally the Category service updates the URL categories accordingly every time the recommendations are being made. Every time a user log into one of the websites using our RS, the website will fetch the recommendations either for an item or for a user.

\subsection{RS Components}
\paragraph{Approximate Nearest Neighbor}
We employed Spotify's Annoy library \cite{li2019approximate} for approximate nearest neighbor (ANN) searches, which proved to be significantly more time-efficient compared to exact nearest neighbor algorithms, such as those provided by the SK-Learn library. A detailed comparison of accuracy and processing time between the exact and approximate models is presented in the Offline Evaluations section under Experiments. In Section \ref{annoy} we will compare the performance of the nearest neighbor and approximate nearest neighbor.

\paragraph{Post/Pre-Filtering}

To enhance recommendation quality, web pages underwent pre- and post-filtering. In the content-based (CB) algorithm, a percentage of low-view pages were excluded to avoid displaying outdated content. After identifying nearest neighbors, pages exceeding a certain cosine similarity threshold were also removed to prevent recommending overly similar content, such as duplicate news coverage from different reporters. For the collaborative filtering (CF) algorithm, users with fewer than two views and pages with fewer than five views were omitted from the matrix to increase density. This adjustment helps reduce inadequate recommendations and addresses the cold-start problem.

\paragraph{Cold Start Fall-back Strategy}
In our algorithms, to mitigate low fill-rate or coverage issues when data on a user or item is insufficient, we employ a self-organized fallback strategy that provides generalized recommendations. If CF recommendations are unavailable, the RS defaults to CB, category-trend, or website-trend suggestions. Additionally, for websites with higher item churn, we run our algorithms more frequently to maintain recommendation relevance.

\section{Dataset}

The data for this study was collected from various websites utilizing our RS. Approximately 70 websites currently implement Yektanet's RS, and we aimed to include a diverse range of platform types and scales in our research. Table 1 presents the number of users and items across each of these websites.

\begin{table}[h]
\centering
\small  
    \begin{tabular}{*{1}{c}|ccp{30mm}}
        \toprule
        
        \bf{Website}&\bf{\# Items}&\bf{\# Users}&\bf{Type of Website}\\
        \midrule
        Tabnakjavan.com&16.4K&8.388M&News\\
        Tapmusics.ir&16.38K&1.048M&Movie Criticize\\
        Paroshat.com&8.19K&4.194M&Movie Criticize\\
        Entekhab.ir&32.67K&2.097M&News\\
        Faradeed.ir&14.2K&1.4M&News\\
        Chetor.com&8.19K&2.1M&Publishing\\
        Academicfiles.ir&624&16.38K&E-commerce\\
        \bottomrule
    \end{tabular}
    
    \caption{Numbers of users and items for each website used in our analysis.}
    \label{tab:useritem}
\end{table}

\section{Manual Analysis}

To interpret the functionality of different sections within the algorithm, we developed a manual analysis approach. A team of recommender system experts, product managers, and domain specialists collaborated to assess the quality of recommendations produced by various algorithms configured with distinct hyper-parameters. Certain hyper-parameters, particularly those affecting semantic similarity, lacked suitable offline tuning metrics and couldn’t be effectively optimized within an online evaluation environment, prompting us to adopt this testing method. Key questions that arose and were addressed through this testing process include: 

\begin{itemize}
    \item How far along the text will the extracted string for our word2vec embedding follow? 
    \item How many clusters are suitable for the category trend algorithm?
    \item Is our self-trained word2vec algorithm a good fit, or are we better off retorting to the Persian version of pretrained models such as Farsi-BERT or fastText, trained respectively by Google and Facebook? 
    \item Between 100 and 300, both of which we tried, which one is a better size for our word2vec embedding?
\end{itemize}

\begin{figure*}
\centering
  \subfloat{
  \includegraphics[width=0.45\linewidth]{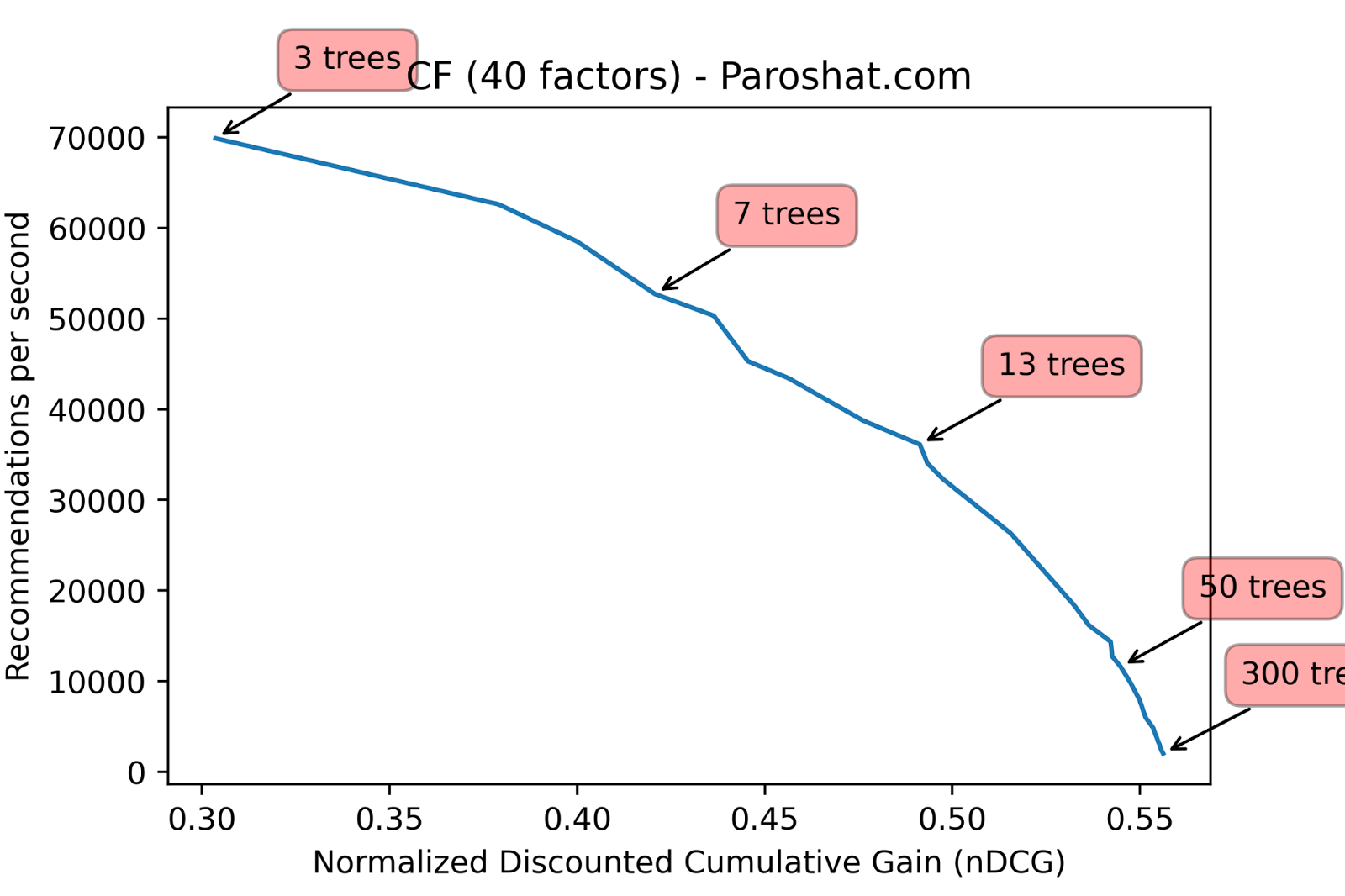}
  }
  \hfill
  \subfloat{
  \includegraphics[width=0.45\linewidth]{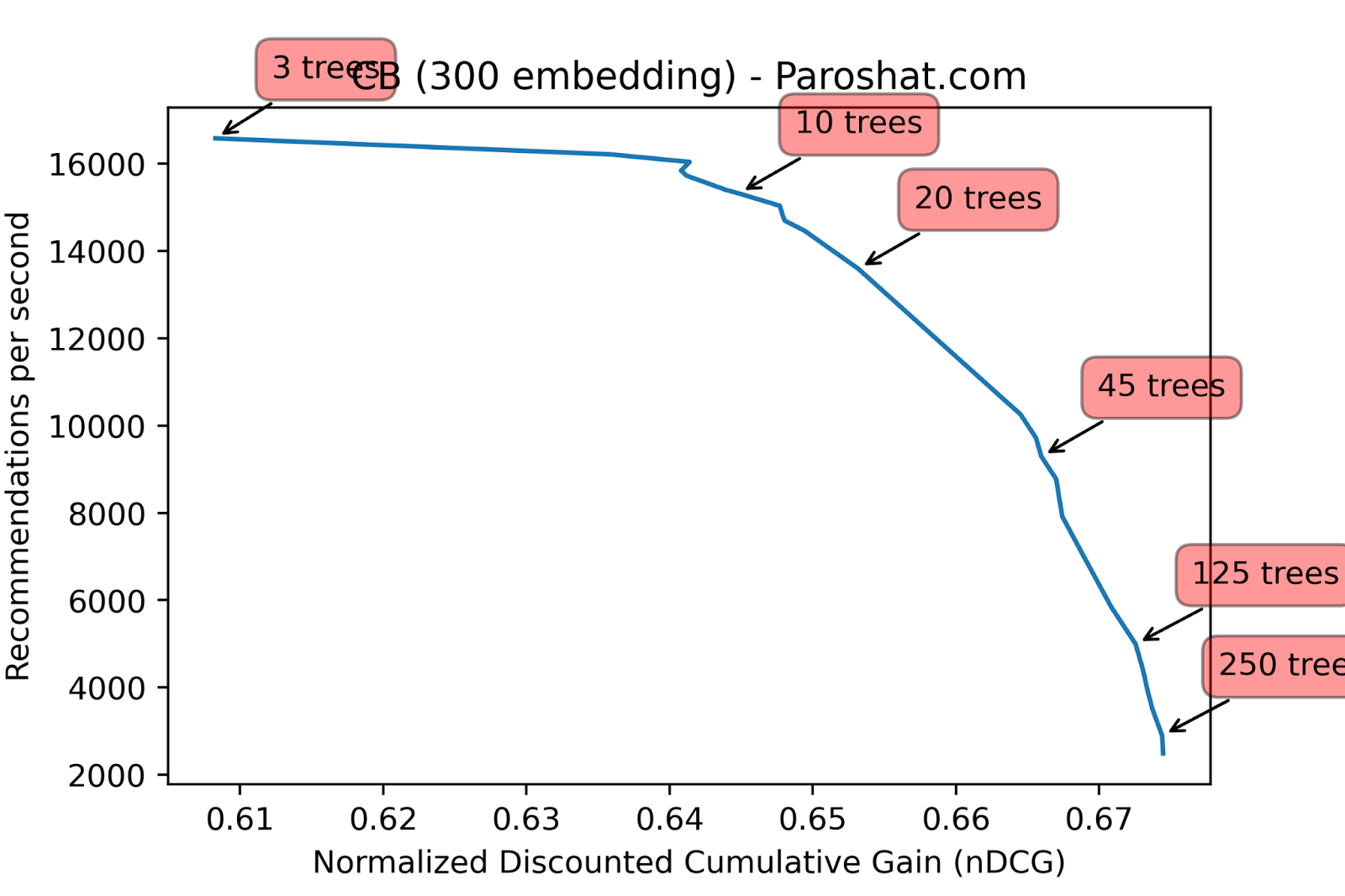}
}

\caption{nDCG with respect to recommendation generation frequency by changing the number of trees of the Annoy library.}
\label{figure1}

\end{figure*}

\begin{figure*}
\centering
  
\includegraphics[width=400pt]{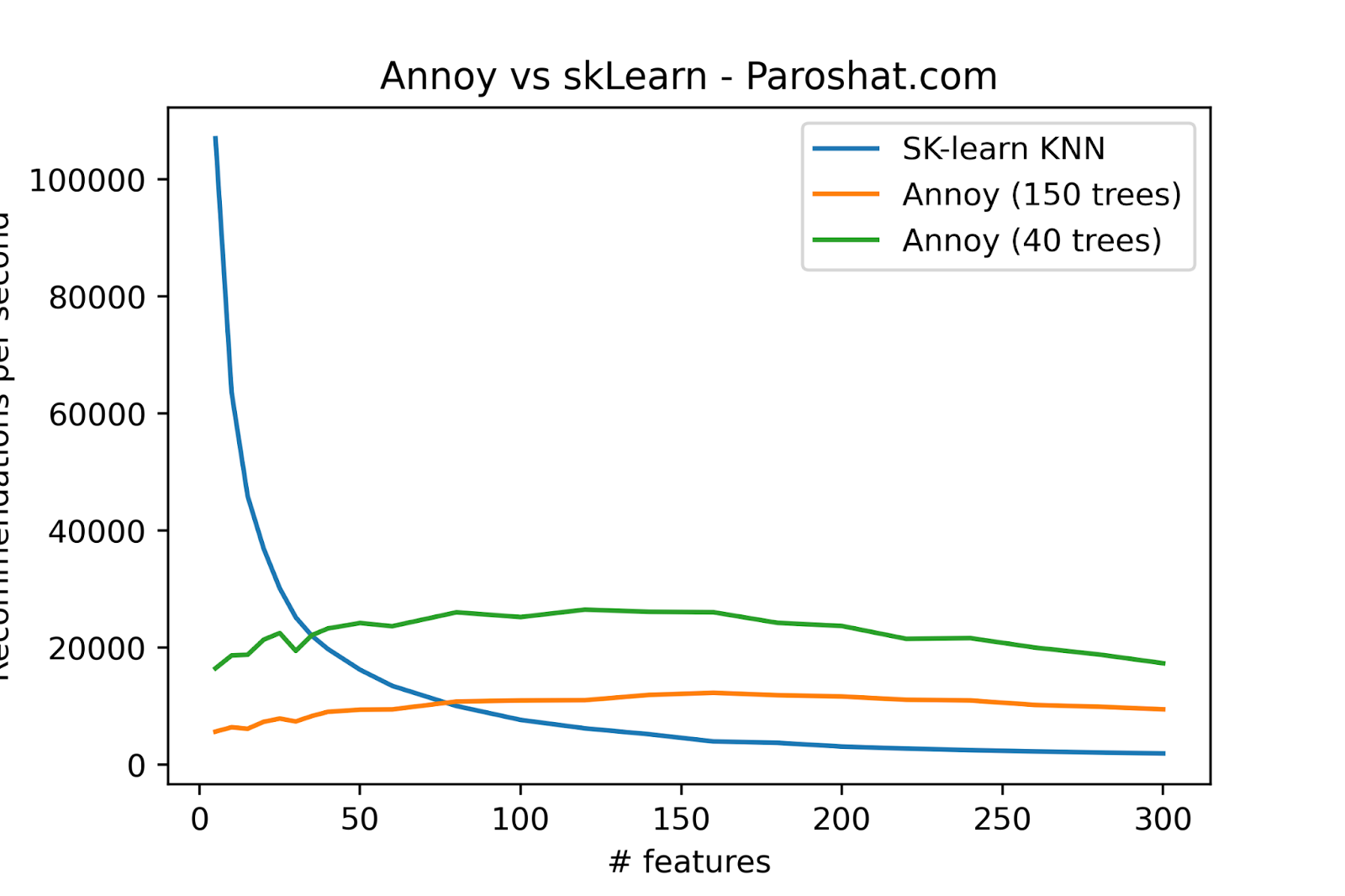}

\caption{Annoy vs SK-Learn on Paroshat.com: The y-axis is the frequency (recommendations generated per second) and the x axis is the number of features in CF algorithm.}
\label{paroshatt}
\end{figure*}

\section{Offline Evaluations}
In our offline tests, we tuned the algorithm hyper-parameters to optimize performance. Specifically, we adjusted parameters in the Annoy library and the Implicit library, used for deploying the approximate nearest neighbor and ALS algorithms, respectively. In the final stage, we examined the impact of popularity bias across different algorithms and refined our hyper-parameters to minimize this effect as much as possible.
\subsection{Metric} 
\paragraph{nDCG} Normalized Discounted Cumulative Gain (nDCG) is a metric used to evaluate the effectiveness of a ranking system. It considers the relevance of the ranked items and their positions in the result set. The nDCG is computed in two steps: first, the Discounted Cumulative Gain (DCG) is calculated using the relevance scores of the items, and then it is normalized by the Ideal Discounted Cumulative Gain (IDCG), which represents the maximum possible DCG for a perfect ranking. The formulation for DCG at rank \( p \) is given by:

$$\text{DCG}_p = \sum_{i=1}^{p} \frac{2^{\text{rel}_i} - 1}{\log_2(i + 1)}$$

where \( \text{rel}_i \) is the relevance score of the item at position \( i \). The nDCG is then computed as:

\[
\text{nDCG}_p = \frac{\text{DCG}_p}{\text{IDCG}_p}
\]

where \( \text{IDCG}_p \) is the DCG for the ideal ranking of the items.

\paragraph{Hit Rate (HR@k)} HR@k is a commonly used metric for evaluating the performance of recommendation algorithms. It measures the proportion of times that a user’s true preference, or "hit," appears in the top \( k \) recommendations generated by the model. Formally, for a user \( u \), a hit occurs if any of the items they have interacted with is present in the top \( k \) items recommended. Let \( R_u \) denote the set of items recommended to user \( u \) and \( T_u \) be the set of items that user \( u \) actually interacted with (e.g., viewed, clicked, or purchased). The Hit Rate@k for user \( u \) is defined as:

\[
\text{HR@k}_u = \begin{cases}
      1 & \text{if } R_u \cap T_u \neq \emptyset, \\
      0 & \text{otherwise}.
   \end{cases}
\]

The overall Hit Rate@k is then calculated as the average hit rate across all users, which can be expressed as:

\[
\text{HR@k} = \frac{1}{|U|} \sum_{u \in U} \text{HR@k}_u,
\]

where \( |U| \) is the total number of users. This metric reflects the effectiveness of the model in providing relevant items within a limited recommendation list. A higher HR@k score indicates that the model is more successful in delivering items of interest to users within the top \( k \) recommended items.

\begin{figure}
\centering
    \subfloat{
  \includegraphics[width=0.88\linewidth]{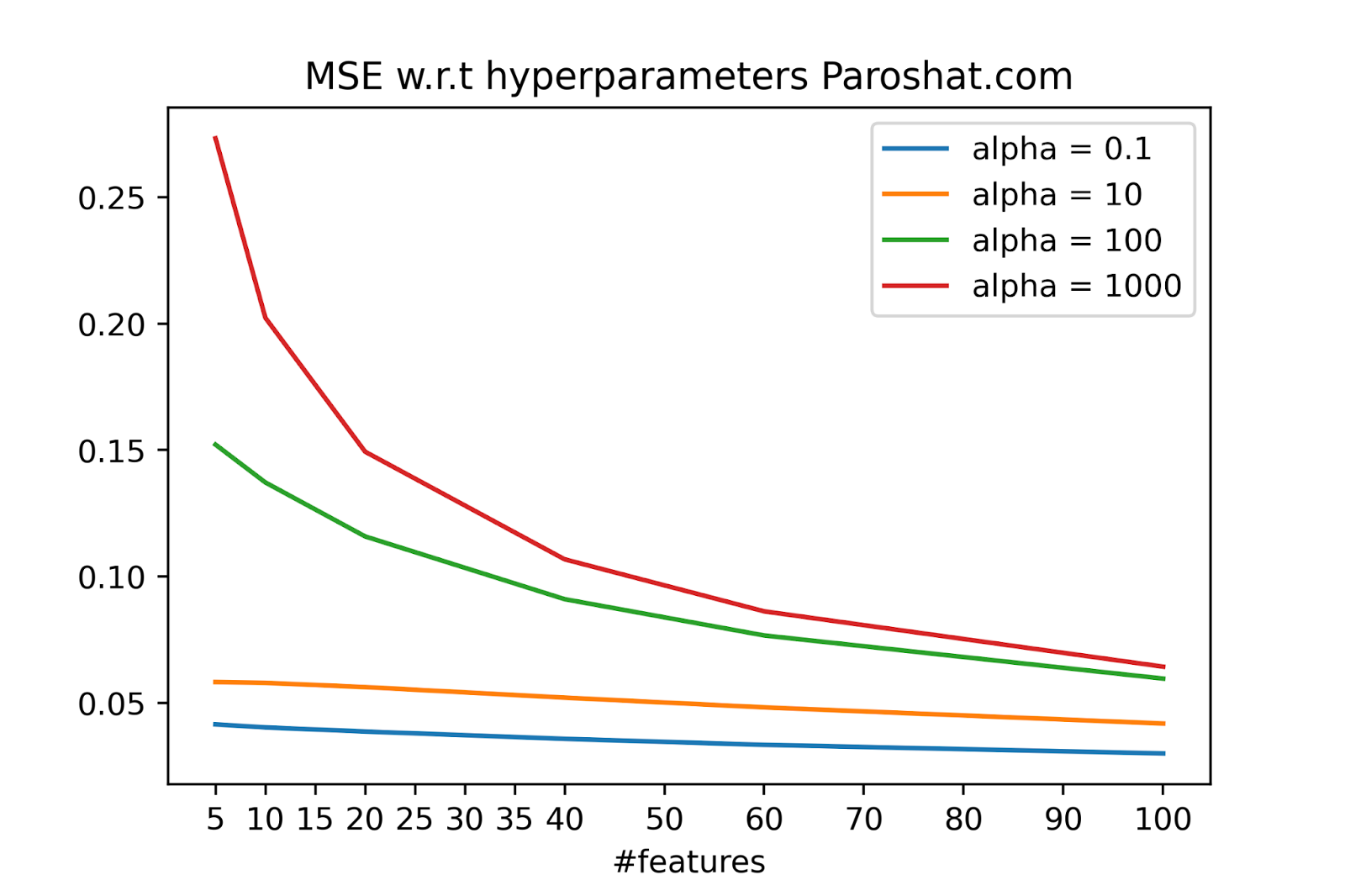}
  }
  \subfloat{
  \includegraphics[width=0.88\linewidth]{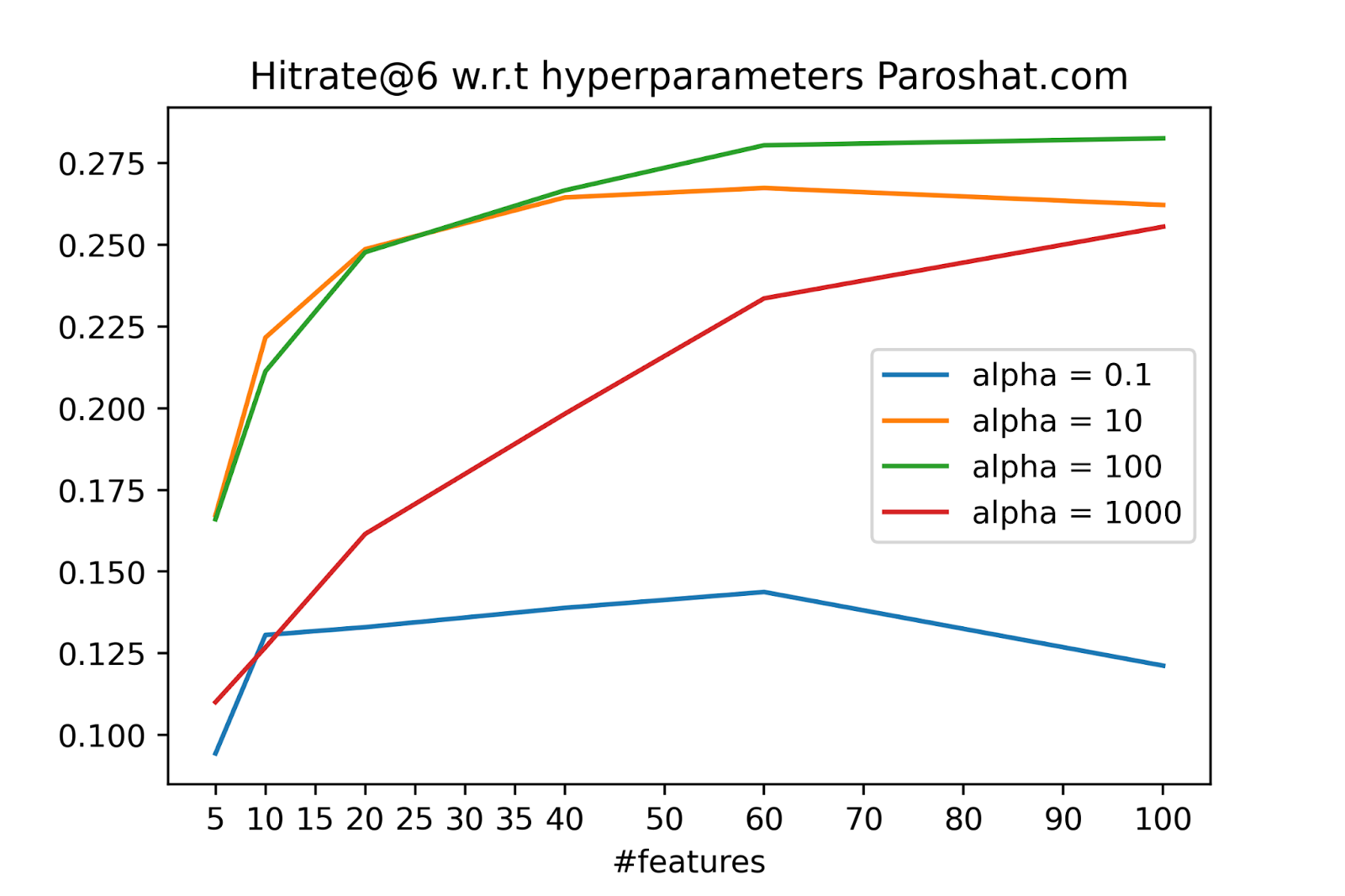}
    }
\caption{Metrics wrt hyper-parameters for ALS Alg.}
\label{fig:mh}
\end{figure}

\subsection{Annoy library}
\label{annoy}
The Annoy library is a C++ library used to search for points in space that are close to a given query point, making it ideal for identifying similar items within our large embedding spaces. These embeddings may be generated through collaborative filtering (CF) or word2vec. We evaluated the Annoy-based approximate nearest neighbor (ANN) algorithm by examining both its time efficiency and accuracy. Accuracy refers to the correlation between items identified by ANN as similar to a primary item and the exact nearest neighbors, using the nDCG ranking measure to assess relevance. Given the large scale of our dataset, time efficiency, or query time, is crucial. We measure throughput as the number of recommendations generated per second within the embedding space.

The Annoy library features two largely independent hyper-parameters: n\_trees and search\_k. We utilized the default setting for search\_k while tuning the n\_trees hyper-parameter based on nDCG and time-throughput metrics. As shown in Figure \ref{figure1}, the nDCG values for most websites do not exceed 0.5, indicating a satisfactory level of similarity, as excessive similarity is not ideal. After evaluating the trade-off between time efficiency and accuracy, we settled on an n\_trees value of 50 for our algorithm. 

Figure \ref{paroshatt} illustrates the impact of embedding feature count (embedding size) on the time performance of nearest neighbor algorithms. A comparison was made between the nearest neighbor algorithms from the SK-Learn library and the Annoy library, using n\_trees values of 50 and 150. The diagram shows that the time performance of the Annoy library improves positively once the feature count exceeds a certain threshold, which is influenced by the number of items in the domain. Based on our findings, we determined that the SK-Learn library is more suitable for websites with fewer items and users, while the Annoy library is better suited for the majority of our publishers, which tend to have larger datasets.

\begin{figure}
\centering
    \subfloat{
    \includegraphics[width=0.7\linewidth]{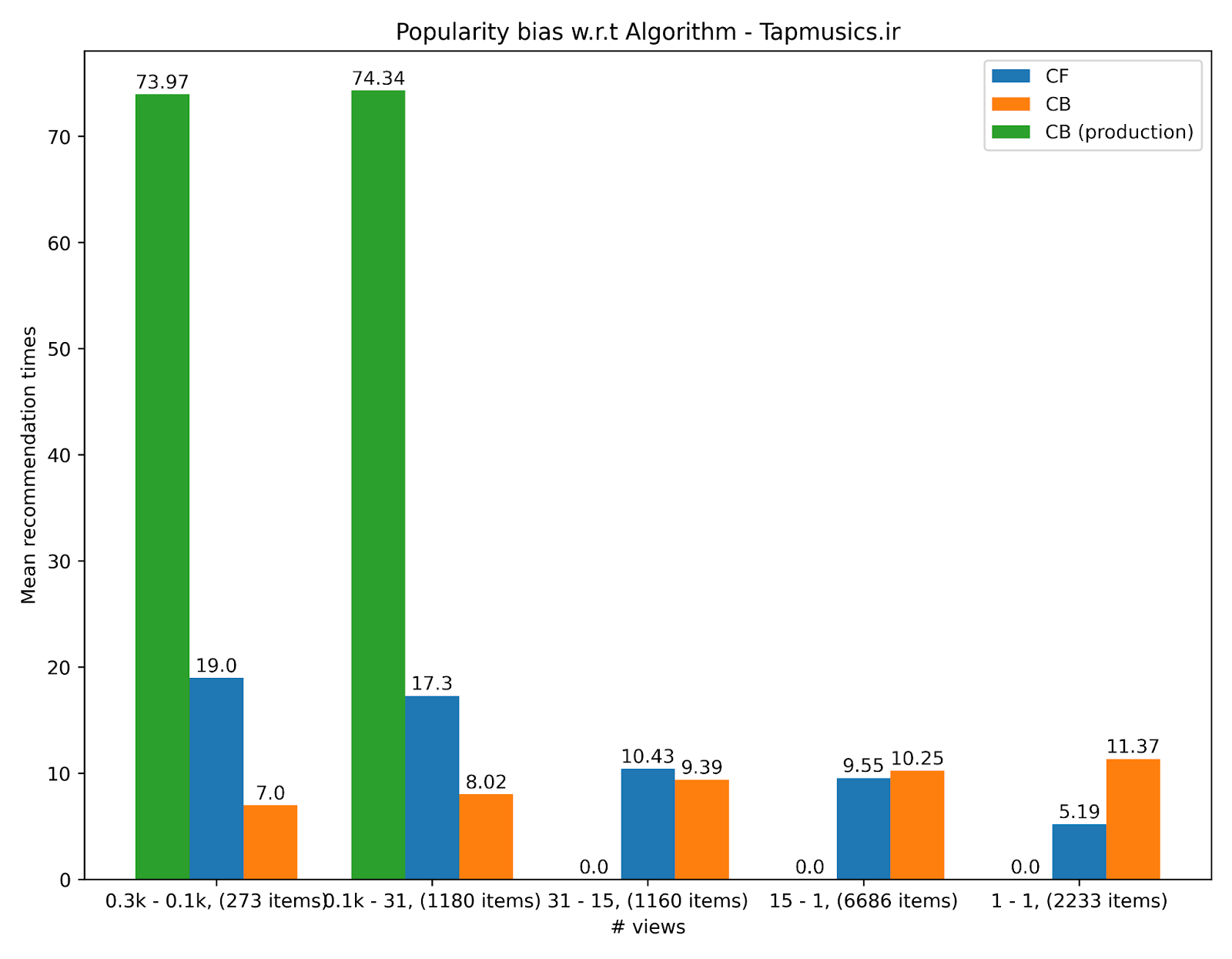}
    }

    \subfloat{
    \includegraphics[width=0.8\linewidth]{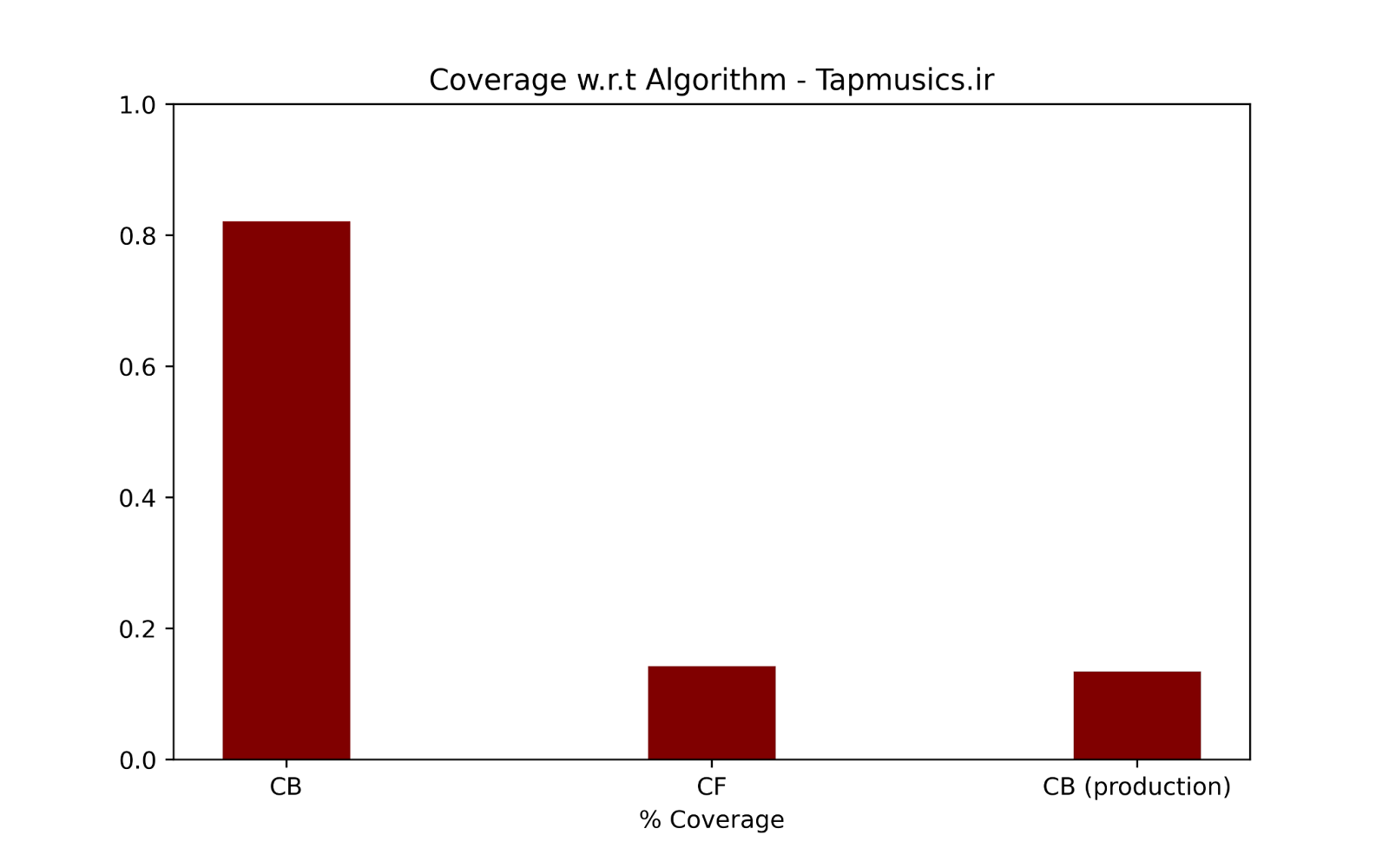}
    }
    \hfill
    \subfloat{
    \includegraphics[width=0.8\linewidth]{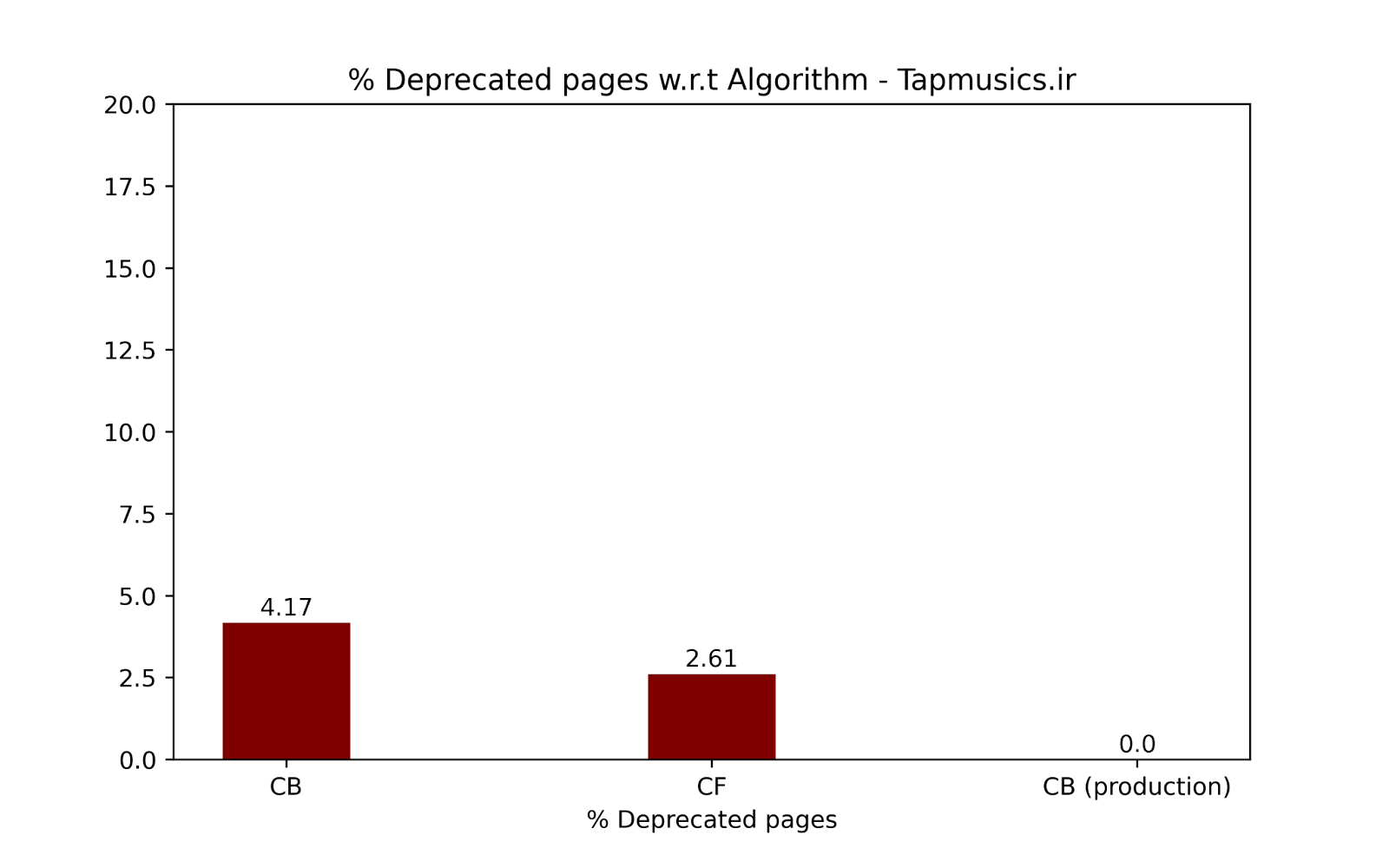}
    }

\caption{Popularity Bias}
\label{pop}
\end{figure}

\begin{figure*}
\centering
    \includegraphics[width=1\textwidth]{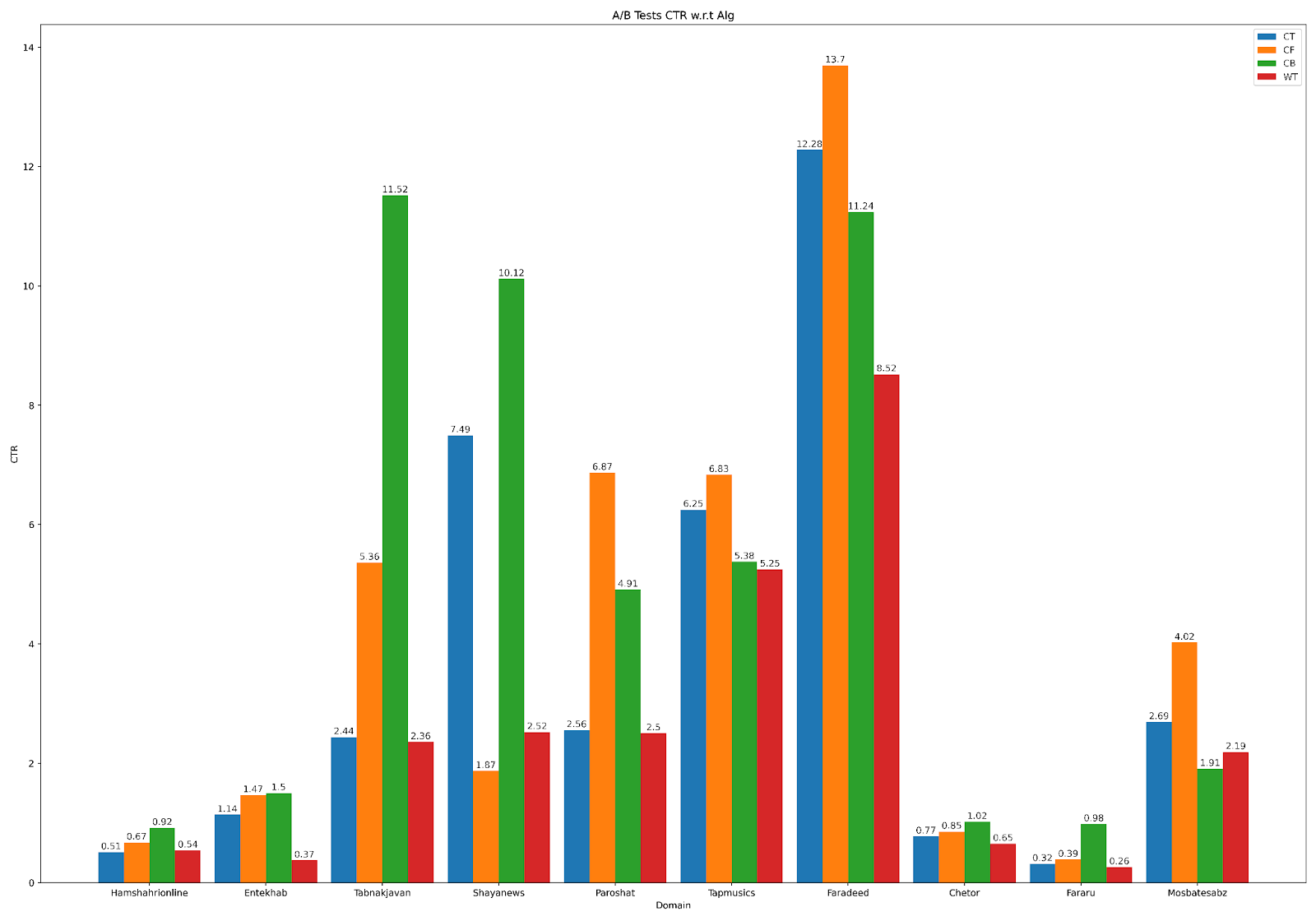}

\caption{CTR Metric per Algorithms}
\label{ctrpalg}
\end{figure*}

\begin{figure*}
\centering
    \includegraphics[width=1\textwidth]{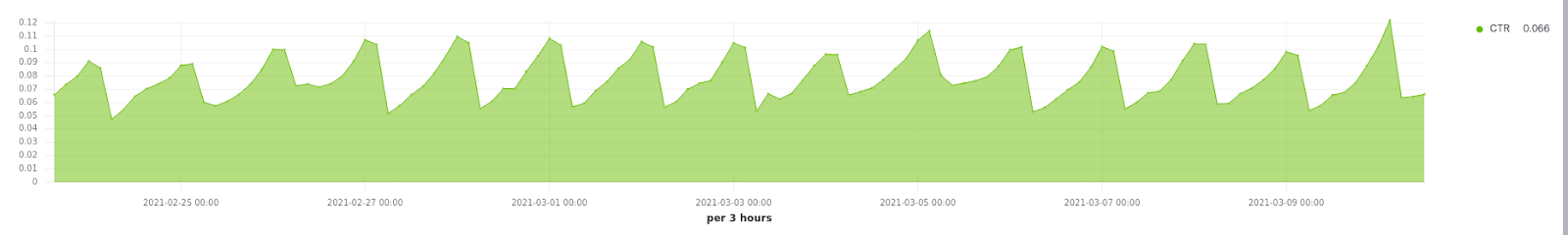}

\caption{System Requests in Two Weeks}
\label{week}
\end{figure*}

\subsection{Alternating Least Square}
\label{als}
The Implicit library has been engaged to deploy the CF algorithm, using Alternative Least Square (ALS) method, which itself adopts the stochastic gradient descent to create user and item matrices. The two main hyper-parameters being alpha and embedding size have been tuned.

Alpha is the strictness of the simulation for the primary matrix; the higher the value of alpha, the more similar the approximate matrix (which is the product of the multiplication of the user matrix by the item matrix) is to the primary matrix. The closer alpha inches to zero, the more random the recommendations would be. Alpha is somewhat a demonstration of recommendation novelty; in higher values, it is more likely for an overfitting issue to arise as most recommended items will be almost identical to the items previously viewed. If the value for alpha is among lower numbers, randomness will have ensued in the algorithm and irrelevant items will also be promoted.

The second hyper-parameter is the number of features (the dimension of embedding space or embedding size). The higher the value for this hyper-parameter, the more the approximate matrix is similar to the primary matrix. In order to improve matrix density, less active users and non-recent items with less views are eliminated in the first stage. This step will be crucial to  mitigate popularity bias. Afterwards, ALS is trained on the matrix.

HR@k criteria were employed for hyper-parameter tuning on train and test data. 20\% of viewed pages were eliminated from the matrix to be used as test data; an attempt was made for the eliminated pages to be among the ones viewed not long ago by users. As observable in Figure 4, hit rate increases both in testing and training with the increase of alpha and number of features until they reach roughly static conditions.

We finally gathered that the value for alpha should be situated somewhere between 60-100 and feature count should be somewhere between 20-40. Outside of these latter limits, the query time computation for Annoy library would take too long and time throughput will decrease, resulting in less authenticity for the algorithm. MSE loss and hit rate for each hyper-parameter appears in Figure \ref{fig:mh}.

\paragraph{Popularity bias}

The CF algorithm is known to be associated with popularity bias problem. As shown in Figure \ref{pop}, the CF algorithm recommends popular items more frequently than non-popular items, while CB algorithms do not segregate non-popular items. A differentiation between items based on popularity is not necessarily undesirable; at very least, it aids with decreasing the prevalence of deprecated item previews. It is recognizable that a filtered CB algorithm does not show deprecated items (items with very few views over time).

We have divided items into groups based on popularity and calculated the average times an item is recommended with each of the three algorithms (CB algorithm, pre-filtered CB algorithm and CF algorithm). This has been done with two different alpha values in the CF algorithm in an attempt to minimize popularity bias. Figure \ref{pop}, alternatively, displays deprecated page recommendations, suggesting that CF algorithms are generally not as prone to displaying deprecated items. A CB algorithm employed on a set of items in which deprecated pages are filtered beforehand, is likewise capable of the same thing.

\begin{figure}
\centering
    \subfloat{
    \includegraphics[width=0.75\linewidth]{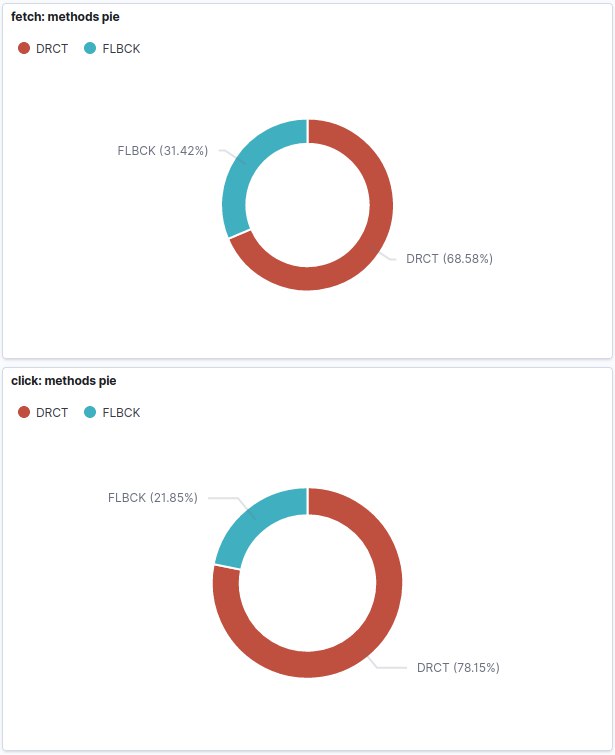}
    }

\caption{Fall-back Statistics}
\label{fallnack}
\end{figure}

\begin{figure}
\centering
    \subfloat{
    \includegraphics[width=0.77\linewidth]{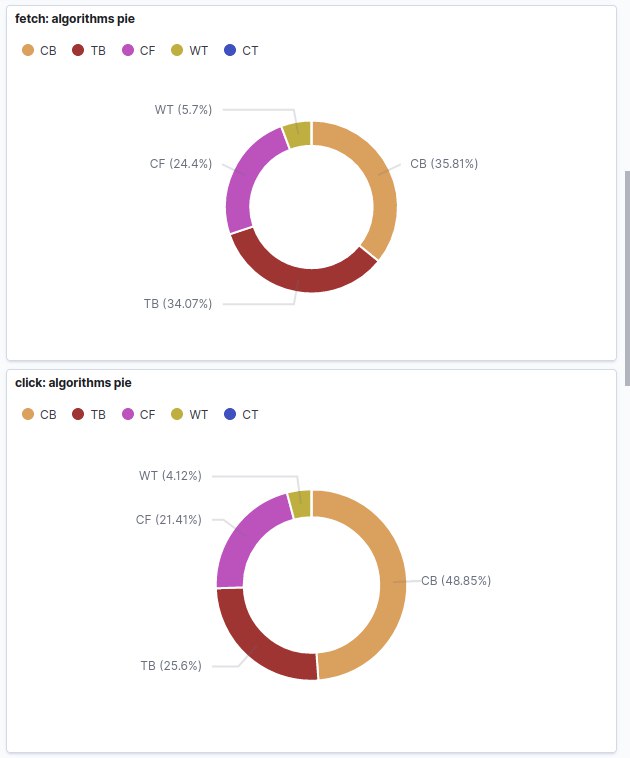}
    }

\caption{Fetch and Click Ratios for Each Algorithms}
\label{algratio}
\end{figure}

\section{Online}

We examined algorithm performance through online metrics such as CTR. These tests were organized in October 2021. Figure \ref{week} shows the CTR of the system over a week; The average CTR for the week is reported 6\%.

Recommendation boxes appear below each item and contain four to six suggestions generated by algorithms like CB, CF, website-trend, category-trend, and user-based models. Our fetch and click services log events whenever a recommendation box is displayed or a suggestion is clicked. These logs are stored in an Elasticsearch database using Fluent Bit, and we use the Elastic Kibana interface to visualize the results. 

In Figure \ref{ctrpalg}, the box plots show CTR for each algorithm, highlighting that hybrid algorithms achieve the highest overall performance. Additionally, CB algorithms perform particularly well on news websites, while CF algorithms are optimal for music and movie platforms.

The CB algorithm successfully generates recommendations for nearly 100\% of pages, while the CF algorithm provides recommendations for only about half of the pages. This difference results from pre-filtering applied before sending pages and users to the CF algorithm, which excludes users with fewer than two views and pages with fewer than five views, creating a denser matrix. The second part of the diagram illustrates fill rates for users, representing the percentage of users for whom the user-based algorithm was able to generate recommendations.

Figure \ref{fallnack} illustrates the system's fallback times. As described, a fallback strategy is used when there are insufficient recommendations for an item or user, prompting the system to switch to more general methods—for example, showing content-based (CB) recommendations instead of collaborative filtering (CF) when CF recommendations are unavailable. The fetch service is invoked when a user navigates to a page, and that page requests recommendations for either the user or the item. Similarly, the click service is activated when a recommendation is clicked, redirecting the user to the targeted page. Approximately 31.4\% of fetch requests end up using this fallback strategy, with 21.85\% of these resulting in a click.

Figure \ref{algratio} displays the percentage of recommendations served by each algorithm. For instance, 31.81\% of the recommendations shown to users were generated by the CB algorithm, while 48.85\% of the clicks were on CB-generated recommendations.

\bibliographystyle{ACM-Reference-Format}
\bibliography{sigconf}

\section{Acknowledgments}

This project was conducted in Iran, at Yektanet Company. All data was gathered with user consent, and website content was crawled following notification of the site owners.

\end{document}